\documentclass[aps,pra,twocolumn,showpacs,preprintnumbers,amsmath,amssymb]{revtex4}
\usepackage{graphicx}
\usepackage{bm}
\usepackage{dcolumn}
\usepackage{xcolor}

\usepackage[breaklinks=true,colorlinks,citecolor=blue,linkcolor=blue,urlcolor=blue]{hyperref}

\def\sc#1#2#3{Science {\bf #1}, #2 (#3)}

\def\rmp#1#2#3{Rev. Mod. Phys. {\bf #1}, #2 (#3)}
\def\prl#1#2#3{Phys. Rev. Lett. {\bf #1}, #2 (#3)}
\def\pra#1#2#3{Phys. Rev. A {\bf #1}, #2 (#3)}

\def\epjd#1#2#3{Eur. Phys. J. D {\bf #1}, #2 (#3)}

\def\pla#1#2#3{Phys. Lett. A {\bf #1}, #2 (#3)}

\def\jltp#1#2#3{J. Low Tem. Phys. {\bf #1}, #2 (#3)}

\def\ejp#1#2#3{Eur. J. Phys. {\bf #1}, #2 (#3)}
\def\pramana#1#2#3{Pramana - J. Phys. {\bf #1}, #2 (#3)}

\def\noi{\noindent}
\def\bc{\begin{center}}
\def\ec{\end{center}}
\newcommand{\bea}{\begin{equation}}
\newcommand{\eea}{\end{equation}\noi}
\newcommand{\ber}{\begin{eqnarray}}
\newcommand{\eer}{\end{eqnarray}\noi}
\begin{document}
\title{Energy fluctuation and discontinuity of specific heat}
\author{Shyamal Biswas$^1$}\email{sbsp [at] uohyd.ac.in}
\author{Joydip Mitra$^{2}$}
\author{Saugata Bhattacharyya$^{3}$}
\affiliation{$^1$School of Physics, University of Hyderabad, C.R. Rao Road, Gachibowli, Hyderabad-500046, India\\
$^{2}$Department of Physics, Scottish Church College, 1 \& 3 Urquhart Square, Kolkata-700006, India\\
$^{3}$Department of Physics, Vidyasagar College, 39 Sankar Ghosh Lane, Kolkata-700006, India}

\date{\today}

\begin{abstract}
Specific heat per particle ($c_v$) of an ideal gas, in many occasions, is interpreted as energy fluctuation per particle ($\triangle\epsilon^2$) of the ideal gas through the relation: $\triangle\epsilon^2=kT^2c_v$, where $k$ 
is the Boltzmann constant and $T$ is the temperature. This relationship is true only in the classical limit, and deviates significantly in the quantum degenerate regime. We have analytically explored quantum to classical 
crossover of this relationship, in particular, for 3-D free Bose and Fermi gases. We also have explored the same for harmonically trapped cases. We have obtained a hump of $\triangle\epsilon^2/kT^2c_v^{(\text{cl})}$ 
around the condensation point for 3-D harmonically trapped Bose gas. We have discussed the possibility of occurring phase transition with discontinuity of heat capacity from existence of such a hump for other Bose and Fermi systems.
\end{abstract}
\maketitle
\section{Introduction}
Ultracold atomic gas in an optical trap is a favourite hunting ground for theoreticians and experimentalists \cite{Pitaevskii,Bloch,Giorgini}. Within the last 
two decades numerous works were done on this topic \cite{Pitaevskii,Bloch,Giorgini}. Quite a few major branches e.g. Bose-Einstein condensation in harmonic trap \cite{Pitaevskii}, ultracold Fermi gas in harmonic trap \cite{Giorgini}, ultracold gas in optical lattice \cite{Bloch}, superfluid-Mott insulator transition in optical lattice \cite{Bloch,Sengupta}, etc have evolved from this single topic showing its potential. However, the thermodynamic properties of the ultracold atomic gases have always been the centre of attraction. Bose-Einstein condensation fraction \cite{Ketterle,Ensher,Griesmaier,Biswas-PLA}, temperature dependence of energy and specific heat of ultracold Fermi gases \cite{Jin,Kinast,Biswas-EPJD}, momentum distribution for harmonically trapped Bose gas \cite{Anderson}, momentum distribution for harmonically trapped Fermi gas \cite{Hulet}, temperature dependence of the chemical potential \cite{Luo}, temperature dependence of critical number of particles for collapse of attractively 
interacting Bose gas \cite{Biswas-2009}, temperature dependence of thermodynamic properties of a unitary Fermi gas \cite{Ku-2012}, etc have already been studied in this regard. As the discontinuity of specific heat (at constant volume and average particle number) confirms about occurrence of the second order phase transition, measurement of the specific heat is very important. Surprisingly, specific heat for harmonically trapped Bose gas, as far as we know, has not yet been precisely measured, specially around the condensation point. This is because of difficulties associated with the inhomogeneity of the trapped condensate, and the corresponding difficulties in analyzing the density profiles within local density approximation \cite{Shiozaki-2014}. If this is so difficult to be measured, is there any other measurable thermodynamic variable which confirms discontinuity of specific heat? This paper finds a possible answer to this question from a general perspective. 

Specific heat is very often interpreted as energy fluctuation. Energy fluctuation, for a single particle subsystem, is nothing but the variance of energy $\triangle\epsilon^2=\bar{\epsilon^2}-\bar{\epsilon}^2$, 
where $\bar{\epsilon}=\sum_i\epsilon_ip_i$, $\bar{\epsilon^2}=\sum_i\epsilon_i^2p_i$, $\epsilon_i$ is the energy eigenvalue of the particle at the $i$th eigenstate, and $p_i$ is the thermal equilibrium probability of the 
$i$th eigenstate. As long as we take this probability to be $p_i=\text{e}^{-\epsilon_i/kT}/Z$, where $Z=\sum_i\text{e}^{-\epsilon_i/kT}$ is the canonical partition function, we obtain the familiar relationship: 
$\triangle\epsilon^2=kT^2c_v^{(\text{cl})}$ where $c_v^{(\text{cl})}$ is the classical specific heat of a single particle. The same probability is eventually applicable for a single particle of an ideal classical gas, 
and as a result, this relationship is true only for a classical gas.

For a quantum (Bose or Fermi) gas, above probability can be given (considering a grand canonical ensemble) by $p_i=\bar{n}_i/N$, where $\bar{n}_i$ is average number of particles occupying the $i$th eigenstate and $N 
(=\sum_{i}\bar{n}_i)$ is the total average number of particles. Average number of particles, occupying the $i$th eigenstate, is, of course, given by the Bose-Einstein ($-$) or Fermi-Dirac ($+$) statistics
\begin{eqnarray}\label{eqn1}
\bar{n}_i=\frac{1}{\text{e}^{(\epsilon_i-\mu)/kT}\mp1}
\end{eqnarray}
where $\mu$ is the chemical potential of the quantum (Bose or Fermi) gas. Difference between the probabilities ($p_i$) of the classical and the quantum gas stems essentially from the nonzero fugacity 
($z=\text{e}^{\mu/kT}$) of the quantum gas. 

Bose-Einstein or Fermi-Dirac statistics as expressed above is applicable only in the thermodynamic limit where microcanonical, canonical and grand canonical ensembles of statistical mechanics reproduce the same result \cite{ensemble, Biswas-PLA}. However, grand canonical ensemble is convenient to work with, in particular, for the quantum (Bose and Fermi) gases.

In the following, we will see how a nonzero fugacity plays a vital role in the deviation of $\triangle\epsilon^2$ from $kT^2c_v^{(\text{cl})}$, in particular, for 3-D free 
\cite{pathria} and harmonically trapped \cite{pitaevskii-book} ideal Bose and Fermi gases.
\section{Energy fluctuations}
\subsection{For free Bose and Fermi gases}
For a free gas, we can replace the single particle energy eigenstates $\{i\}$ by the single particle momenta $\{\textbf{p}\}$ so that $\epsilon_i\rightarrow\epsilon_\textbf{p}=p^2/2m$ where $m$ is the mass of the single 
particle. Degeneracy of this level, within the semiclassical approximation, is given by $\frac{V4\pi p^2dp}{(2\pi\hbar)^3}$, where $V$ is the volume of the system. In the thermodynamic limit ($N\rightarrow\infty,
V\rightarrow\infty, N/V=\bar{n}=const.$), straightforward textbook level calculation, for the energy fluctuation ($\triangle\epsilon^2=\sum_i\epsilon_i^2[\bar{n}_i/N]-(\sum_i\epsilon_i[\bar{n}_i/N])^2=\frac{V4\pi}{N(2\pi\hbar)^3}
\int_0^\infty\epsilon_{\textbf{p}}^2\bar{n}_{\textbf{p}}p^2\text{d}p-\frac{V^2(4\pi)^2}{N^2(2\pi\hbar)^6}[\int_0^\infty\epsilon_{\textbf{p}}\bar{n}_{\textbf{p}}p^2\text{d}p]^2$) of a 3-D free Bose gas, results in
\begin{eqnarray}\label{eqn2}
\triangle\epsilon^2=\bigg\{\begin{matrix}&(kT)^2\big[\frac{15}{4}\frac{\zeta(\frac{7}{2})}{\zeta(\frac{3}{2})}\big(\frac{T}{T_c}\big)^{\frac{3}{2}}-\frac{9}{4}\frac{\zeta^2(\frac{5}{2})}{\zeta^2
(\frac{3}{2})}\big(\frac{T}{T_c}\big)^{3}]~\text{for}~\frac{T}{T_c}\le 1\\
&(kT)^2\big[\frac{15}{4}\frac{\text{Li}_{7/2}(z)}{\text{Li}_{3/2}(z)}-\frac{9}{4}\frac{\text{Li}_{5/2}^2(z)}{\text{Li}_{3/2}^2(z)}\big] \ \ \ \ \ \ \ \ \ ~\text{for}~\frac{T}{T_c}> 1
\end{matrix}
\end{eqnarray}
where $T_c=\frac{2\pi\hbar^2}{mk}\big(\frac{\bar{n}}{\zeta(3/2)}\big)^{2/3}$ \cite{pathria} is the Bose-Einstein condensation temperature and $\text{Li}_j(z)=z+\frac{z^2}{2^j}+\frac{z^3}{3^j}+...$ is a polylog function of
 order $j$. Similar straightforward calculation, for a 3-D free Fermi gas, results in
\begin{eqnarray}\label{eqn3}
\triangle\epsilon^2=(kT)^2\bigg[\frac{15}{4}\frac{\text{Li}_{7/2}(-z)}{\text{Li}_{3/2}(-z)}-\frac{9}{4}\frac{\text{Li}_{5/2}^2(-z)}{\text{Li}_{3/2}^2(-z)}\bigg].
\end{eqnarray}
Comparing the above results with the textbook results for $kT^2$ times $c_v$ \cite{pathria}, one can easily check that $\triangle\epsilon^2\neq kT^2c_v$ except in the classical limit ($z\rightarrow0$). To visualize how the above 
results deviate from their unique classical value ($\frac{3}{2}(kT)^2$), we can plot the right hand sides of Eqns.(\ref{eqn2}) and (\ref{eqn3}) in units of $\frac{3}{2}(kT)^2$ for the entire range of temperature. But, plotting of the 
same is not an easy job until one manages to get temperature dependence of $z$ or of $\mu$ from the implicit relations \cite{pathria}
\begin{eqnarray}\label{eqn4}
\bigg(\frac{\text{Li}_{3/2}(z)}{\zeta(3/2)}\bigg)^{2/3}=\frac{T_c}{T}
\end{eqnarray}
for 3-D free Bose gas, and \cite{pathria}
\begin{eqnarray}\label{eqn5}
\big(-\text{Li}_{3/2}(-z)\Gamma(5/2)\big)^{2/3}=\frac{T_F}{T}
\end{eqnarray}
for 3-D free Fermi gas whose Fermi temperature is given by $T_F=\frac{\hbar^2}{2mk}(6\pi^2\bar{n})^{2/3}$. Approximate analyses of the temperature dependence of the chemical potentials were done by Biswas \textit{et al} from the above 
two equations \cite{mu}. They obtained temperature dependent approximate formulas of $\mu$ not only for the free Bose and Fermi gases, but also for the harmonically trapped Bose and Fermi gases. Using their appropriate formulas 
of $\mu$, we plot $\triangle\epsilon^2$s in units of $\frac{3}{2}(kT)^2$ in FIG.1a for the free Bose gas, and in FIG. 1b for the free Fermi gas.

\subsection{For harmonically trapped Bose and Fermi gases}
For harmonically trapped case, all the particles are 3-D harmonic oscillators, and the energy levels are given by $\epsilon_i=(\frac{3}{2}+i)\hbar\omega$, where $\omega$ is the angular frequency of oscillations. Although the 
degeneracy ($g_i$) of this level is $i^2/2+3i/2+1$ \cite{powell}, yet in the thermodynamic limit ($N\rightarrow\infty$, $\omega\rightarrow0$ \& $N\omega^3=const.$), only the first term of the degeneracy contributes significantly. 
In this limit, the zero point energy can also be neglected. Now, straightforward textbook level calculation, for the energy fluctuation 
($\triangle\epsilon^2=\frac{1}{N}\int_0^\infty\epsilon^2\bar{n}(\epsilon)g(\epsilon)\text{d}\epsilon-\frac{1}{N^2}[\int_0^\infty\epsilon\bar{n}(\epsilon)g(\epsilon)\text{d}\epsilon]^2$) where 
$g(\epsilon)=\epsilon^2/2$ in the thermodynamic limit of a 3-D harmonically trapped Bose gas, results in
\begin{eqnarray}\label{eqn6}
\triangle\epsilon^2=\bigg\{\begin{matrix}&(kT)^2\big[12\frac{\zeta(5)}{\zeta(3)}\big(\frac{T}{T_c}\big)^{3}-9\frac{\zeta^2(4)}{\zeta^2(3)}\big(\frac{T}{T_c}\big)^{6}] \ \text{for}\ \frac{T}{T_c}\le1\\
&(kT)^2\big[12\frac{\text{Li}_{5}(z)}{\text{Li}_{3}(z)}-9\frac{\text{Li}_{4}^2(z)}{\text{Li}_{3}^2(z)}\big] \ \ \ \ \ \ \ \ \ \ \ \ ~\text{for}\ \frac{T}{T_c}>1
\end{matrix}
\end{eqnarray}
where $T_c=\frac{\hbar\omega}{k}\big(\frac{N}{\zeta(3)}\big)^{1/3}$ \cite{pitaevskii-book} is the Bose-Einstein condensation temperature for the trapped system. Similar straightforward calculation, for a 3-D harmonically trapped 
Fermi gas, results in
\begin{eqnarray}\label{eqn7}
\triangle\epsilon^2=(kT)^2\bigg[12\frac{\text{Li}_{5}(-z)}{\text{Li}_{3}(-z)}-9\frac{\text{Li}_{4}^2(-z)}{\text{Li}_{3}^2(-z)}\bigg].
\end{eqnarray}
Once again, plotting $\triangle\epsilon^2$, for the entire range of temperature, is not an easy job until one manages to get the temperature dependence of $z$ or of $\mu$ from the implicit relations
\begin{eqnarray}\label{eqn8}
\bigg(\frac{\text{Li}_{3}(z)}{\zeta(3)}\bigg)^{1/3}=\frac{T_c}{T}
\end{eqnarray}
for 3-D harmonically trapped ideal Bose gas \cite{pitaevskii-book}, and
\begin{eqnarray}\label{eqn9}
\big(-\text{Li}_{3}(-z)\Gamma(4)\big)^{1/3}=\frac{T_F}{T}
\end{eqnarray}
for 3-D harmonically trapped ideal Fermi gas whose Fermi temperature is given by $T_F=\frac{\hbar\omega}{k}\big(\Gamma(4)N\big)^{1/3}$ \cite{butts}. Biswas \textit{et al} obtained approximate temperature dependent formulas of $\mu$ 
from the above two equations \cite{mu}. Using their approximate formulas of $\mu$, we plot $\triangle\epsilon^2$s in units of $3(kT)^2$ in FIG 1c for the harmonically trapped Bose gas, and in FIG 1d for the harmonically 
trapped Fermi gas.

\begin{figure}
\includegraphics{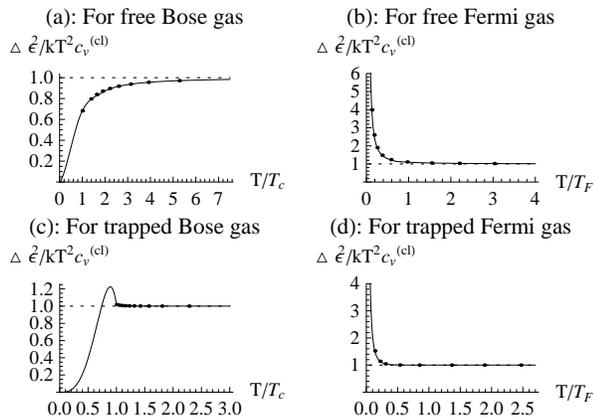}
\caption{Solid lines in FIGs 1a, 1b, 1c, and 1d  (corresponding to the approximate chemical potentials in Ref.\cite{mu}) respectively represent right hand sides of Eqns.(\ref{eqn2}) \& (\ref{eqn3}) in units of 
$\frac{3}{2}(kT)^2$, and Eqns. (\ref{eqn6}) \& (\ref{eqn7}) in units of $3(kT)^2$. Dotted lines represent classical results. Points represent exact graphical solutions.}
\end{figure}

\section{Existence of `humps' in Bose systems}
It is interesting to note in FIG. 1c that $\triangle\epsilon^2/kT^2c_v^{(\text{cl})}$ of the 3-D harmonically trapped ideal Bose gas is smooth unlike its $c_v$, and has a hump just below the condensation point. Maximum in the hump physically means maximum average deviation of the energy of the system from its average energy both in units of average classical energy. Appearance of such a hump over the classical limit may indicate a discontinuity of $c_v$ as well as a phase transition around the condensation point ($T_c$). Here, by the phrase ``over the classical limit'' we mean the average deviation of energy is more than the classical average energy divided by square root of the classical specific heat in unit of the Boltzmann constant.

To measure the energy fluctuation, we need to observe energy distribution for different temperatures. From this observation, both the average of energy and the dispersion of energy can be easily obtained. Temperature can be measured from the the width of the distribution. Thus one can have experimental data for energy fluctuation (dispersion of the energy distribution) for different temperatures. Momentum distribution for the harmonically trapped (Bose/Fermi) gas has already been observed by applying the technique of releasing the harmonic trap \cite{Anderson, Regal}. Energy for the harmonically trapped gas has also been measured by applying this method \cite{Ensher,Jin,Kinast}. Although the specific heat is very difficult to be measured for the 3-D harmonically trapped Bose gas \cite{Shiozaki-2014}, energy fluctuation can be easily measured from energy distribution data by applying the technique of releasing the harmonic trap. 

\begin{figure}
\includegraphics{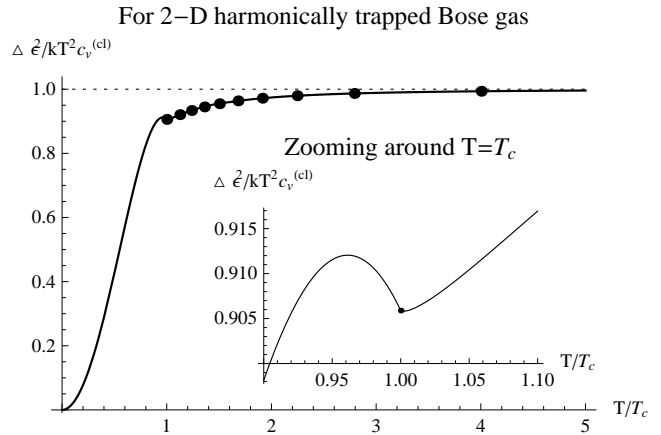}
\caption{Solid line represents $\triangle\epsilon^2$ in units of $2(kT)^2$. Dotted lines represent classical results. Points represent exact graphical solutions.}
\end{figure}

That, appearance of a hump in $\triangle\epsilon^2/kT^2c_v^{(\text{cl})}$ over its classical limit may indicate a discontinuity of $c_v$, is our conjecture.

Bose-Einstein condensation is possible for 3-D free Bose gas. But, there is no discontinuity in its specific heat \cite{pitaevskii-book}. It is clear from FIG. 1 (a) that, for 3-D free Bose gas, there is no hump in $\triangle\epsilon^2/kT^2c_v^{(\text{cl})}$. There is no question of phase transition as well as discontinuity of $c_v$ for an ideal (or weakly interacting) Fermi gas. Obviously, there is no hump in $\triangle\epsilon^2/kT^2c_v^{(\text{cl})}$ as clear in FIGs 1b and 1d for the ideal Fermi gases.

Although Bose-Einstein condensation is possible for 2-D harmonically trapped ideal Bose gas, there is no discontinuity of its specific heat \cite{pitaevskii-book}. It is clear from FIG. 2 that, for the 2-D harmonically trapped ideal Bose gas, although there is a hump in $\triangle\epsilon^2/kT^2c_v^{(\text{cl})}$, it is not over the classical limit.

For the $\lambda$ transition of $^4$He, discontinuity of specific heat was observed way back in 1935 \cite{Keesom}. Theoretical explanation of such discontinuity deals with phonon dispersion which effectively simplifies the interacting Bose system to a 3-D harmonically trapped ideal Bose gas \cite{Feynman}. Considering this simplification, one can easily check that a hump over the classical limit must exist in $\triangle\epsilon^2/kT^2c_v^{(\text{cl})}$ for the liquid He-4 below the $\lambda$ point.

Thus we can verify our conjecture for a certain class of Bose and Fermi systems.

\begin{figure}
\includegraphics{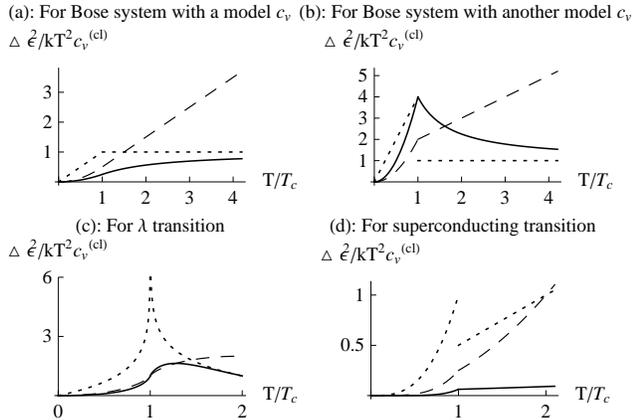}
\caption{The dotted lines in FIGs 3a, 3b, 3c, and 3d represent model specific heats in units of their respective classical values ($c_v/c_v^{\text{(cl)}}$) with respect to scaled temperature ($T/T_c$). Dashed and solid lines respectively represent scaled energy per particle ($\bar{\epsilon}/\sqrt{kT_c^2c_v^{\text{(cl)}}}$) and scaled energy fluctuation ($\triangle\epsilon^2/kT^2c_v^{(\text{cl})}$) corresponding to the respective model specific heats.}
\end{figure}

Our conjecture linking the maximum in the hump of the scaled energy fluctuation ($\triangle\epsilon^2/kT^2c_v^{(\text{cl})}$) and the possible discontinuity of $c_v$ (in unit of $c_v^{\text{(cl)}}$) can be justified by elaboration through a set of model specific heats as shown in FIG. 3 for other Bose and Fermi systems. The model specific heat per particle ($c_v$), as represented by the dotted line in FIG. 3 (a), is appropriate for a homogeneous Bose system \cite{pathria}. We represent average energy per particle ($\bar{\epsilon}$) and approximate energy fluctuation per particle ($\triangle\epsilon^2$) in this figure by the dashed and the solid lines respectively. We obtain the average energy by integrating the model specific heat with respect to the temperature ($\bar{\epsilon}=\int c_v\text{d}T$), and the scaled energy fluctuation by simply squaring it ($\triangle\epsilon^2=\bar{\epsilon^2}-\bar{\epsilon}^2\approx k\bar{\epsilon}^2/c_v^{(\text{cl})}$) so that the approximate 
energy fluctuation becomes exact in the classical limit. FIG. 3 (a) represents a special case of Bose system with no discontinuity in specific heat at $T=T_c$. We do similar exercise in FIG. 3 (b), (c), (d) with different model specific heats with finite discontinuity (e.g. in 3-D harmonically trapped Bose gas) at $T_c$ \cite{Pitaevskii}, with infinite (logarithmic divergence) discontinuity (e.g. in the $\lambda$ transition of $^4$He) at $T_c$ \cite{Feynman}, and with a different finite discontinuity (e.g. in the superconducting/superfluid transition of a strongly interacting Fermi system) \cite{Ku-2012}, respectively. The humps of $\triangle\epsilon^2/kT^2c_v^{(\text{cl})}$s over their respective classical limits and the discontinuities of $c_v$s are simultaneously present only in FIG. 3 (b) and (c) which are appropriate for Bose systems. Although there is a discontinuity of $c_v$ in figure FIG. 3 (d) which is appropriate for a strongly interacting Fermi gas, there is no hump in $\triangle\epsilon^2/kT^2c_
v^{(\text{cl})}$. From these figures we can also essentially verify our conjecture, and say that, for any thermodynamic system, maximum in the hump of $\triangle\epsilon^2/kT^2c_v^{(\text{cl})}$ over its classical limit corresponds to a finite or infinite discontinuity of its specific heat $c_v$.

\section{Conclusion}
Here we have analytically explored quantum to classical crossover in energy fluctuations mainly for free and harmonically trapped quantum gases. We have illustrated how specific heats differ from energy fluctuations for the entire range of temperature. In this regard, we have conjectured about possibility of discontinuity of specific heat in terms of scaled energy fluctuation. 

It is impossible to get exact temperature dependent formulas of chemical potentials for the four systems (3-D free \& harmonically trapped ideal Bose and Fermi gases), as because inverses of polylog functions in Eqn.(\ref{eqn4}), (\ref{eqn5}), (\ref{eqn8}) and (\ref{eqn9}) do not exist in closed forms. For this reason, we have used approximate temperature dependent formulas of chemical potentials from Ref.\cite{mu}. Consequently, plots of $\triangle\epsilon^2$s in FIG 1, become approximate except in the region $0\le T<T_c$ for the Bose gases. To show how good our approximate results are, we have compared them with their exact graphical solutions in the same figures. In the same figures, we also have plotted the classical results just to show the deviations of $\triangle\epsilon^2$s from $kT^2c_v^{\text{(cl)}}$s. These deviations clearly illustrate quantum to classical crossover in energy fluctuation ($\triangle\epsilon^2$) of the four systems of our main interest.

Although the approximate values of $\triangle\epsilon^2$s, for the Bose and Fermi systems, match well with their exact graphical solutions yet it should be mentioned, that, for the Fermi systems, plotting of $\triangle\epsilon^2$s had difficulties, specially in the low temperature regime, for rapidly oscillatory nature of polylog functions in them. For this reason, we have plotted the Sommerfeld's asymptotic forms in FIGs 1b and 1d only for $T/T_F\le0.2$ 
\cite{pathria,asymptotic}.

It is interesting to note that the scaled energy fluctuation ($\triangle\epsilon^2/kT^2c_v^{(\text{cl})}$) of the 3-D harmonically trapped Bose gas has a hump just below $T_c$. Appearance of such a hump over the classical limit may indicate a discontinuity of $c_v$ as well as a phase transition around the condensation point. That, appearance of a hump in $\triangle\epsilon^2/kT^2c_v^{(\text{cl})}$ over its classical limit might point to a discontinuity of $c_v$, is our conjecture for any thermodynamic system. We already verified this conjecture for a number Bose systems. For ideal Fermi gases, as shown in FIG. 1 (b) and 1 (d), existence of the hump and discontinuity of specific heat is impossible. On the contrary, such possibilities exist for a number of interacting or noninteracting Bose gases. Although the specific heat is very difficult to be observed for the 3-D harmonically trapped Bose gas \cite{Shiozaki-2014}, energy fluctuation can be easily measured apparently by applying the technique of releasing 
the harmonic trap \cite{Anderson,Ketterle2}. Thus one can predict the discontinuity of specific heat from the existence of the hump over the classical limit.  

By $c_v$, we not only mean specific heat at constant (effective \cite{Biswas-Pramana} or exact) volume ($V$) but also at constant (average or exact) number of particles $N$ \footnote{Here ``average number of particles'' is applicable for grand canonical ensemble, and ``exact number of particles'' is applicable for microcanonical and canonical ensembles. Specific heat per particle is unique for all the ensembles in the thermodynamic limit.}. It should be mentioned, that, volume of a harmonically trapped system cannot be precisely defined although it can be effectively defined for ultracold situation as $V_{\text{eff}}\sim(\hbar/m\omega)^{3/2}$ \cite{Biswas-Pramana}. For this reason, $c_v$, for trapped system, is very often denoted by $c_N$.

In reality, harmonic traps are not isotropic. In that case, the angular frequency ($\omega$) in all our results should be replaced by the geometric mean of $\omega_x$, $\omega_y$ and $\omega_z$. We have not considered spin degeneracy of particles. Consideration of the same would not change any of our results as because they are intensive variables.

The relation between energy fluctuation and specific heat is a generic property of microcanonical, canonical and grand canonical ensembles of statistical mechanics only in the thermodynamic limit which has been taken into consideration for the entire calculation of this paper \cite{ensemble}. That is why we have expressed both the energy fluctuation and the specific heat in per-particle form. For a finite system, this relationship must be different for different ensembles \cite{ensemble}. Extension of our result for finite system of different statistical ensembles would be an interesting problem.  

To conclude, this paper illustrates quantum to classical crossover in energy fluctuations for a number of Bose and Fermi systems. Appearance of the hump in the scaled energy fluctuation ($\triangle\epsilon^2/kT^2c_v^{(\text{cl})}$) over its classical limit may indicate a discontinuity of $c_v$, is our conjecture for any (Bose or Fermi) system. Inverse of this statement may not always be true as depicted in Fig. 3 (d). All the calculations in this article have been done keeping in mind the scope of general science readers. Extension of our calculations for other spatial dimensions and interacting cases is pretty straightforward and can be taken up by the interested readers. They can verify our conjecture for those cases. Proving our conjecture of course, is an entirely nontrivial and open problem.

\section*{Acknowledgment}
S. Biswas acknowledges financial support form the Department of Science and Technology [DST], Govt. of India under the INSPIRE Faculty Award Scheme [No. IFA-13 PH-70]. For an initial part of this work, S. Biswas further acknowledges hospitality of the Department of Physics, University of Calcutta and financial support form the University Grants Commission [UGC] under the DSKPDF Scheme. We also acknowledge useful discussions with J.K. Bhattacharjee, HRI, India during FIP-2014 conference at the University of Hyderabad. We thank the reviewer for his/her thorough review and highly appreciate all the comments and suggestions, which significantly contributed to improving the quality of the paper.


\begin{thebibliography}{99}
\bibitem{asymptotic}Sommerfeld's asymptotic form of $\triangle\epsilon^2$s in units of $(kT_F)^2$ are $\frac{12}{175}+\frac{\pi^2}{5}(T/T_F)^2+...$ for the free Fermi gas, and $\frac{3}{80}+\frac{\pi^2}{4}(T/T_F)^2+...$ for the harmonically trapped Fermi 
gas. See Ref. \cite{pathria} for Sommerfeld expansion.
\end{thebibliography}
\end{document}